\documentclass{Interspeech}

\title{A Semantic Information-based Hierarchical Speech Enhancement Method Using Factorized Codec and Diffusion Model}

\author[affiliation={1,2}]{Yang}{Xiang} 
\author[affiliation={1}]{Canan}{Huang}
\author[affiliation={1}]{Desheng}{Hu}
\author[affiliation={1}]{Jingguang}{Tian}
\author[affiliation={1}]{Xinhui}{Hu}
\author[affiliation={2}]{Chao}{Zhang}

\affiliation{}{Hithink Research}{Zhejiang, China}
\affiliation{Electrical Engineering Department}{Tsinghua University}{Beijing, China}
\email{\{xiangyang2, tianjingguang, huxinhui\}@myhexin.com, cz277@tsinghua.edu.cn}

\usepackage{comment}
\usepackage{cite}

\usepackage{booktabs}
\usepackage{graphicx}
\usepackage{multirow}
\usepackage{stfloats}
\usepackage{array} 
\usepackage{caption}
\usepackage{amsmath}

\begin{document}

\maketitle

\begin{abstract}
Most current speech enhancement (SE) methods recover clean speech from noisy inputs by directly estimating time-frequency masks or spectrums. However, these approaches often neglect the distinct attributes, such as semantic content and acoustic details, inherent in speech signals, which can hinder performance in downstream tasks. Moreover, their effectiveness tends to degrade in complex acoustic environments. To overcome these challenges, we propose a novel, semantic information-based, step-by-step factorized SE method using factorized codec and diffusion model. Unlike traditional SE methods, our hierarchical modeling of semantic and acoustic attributes enables more robust clean speech recovery, particularly in challenging acoustic scenarios. Moreover, this method offers further advantages for downstream TTS tasks. Experimental results demonstrate that our algorithm not only outperforms SOTA baselines in terms of speech quality but also enhances TTS performance in noisy environments.

\end{abstract}

\section{Introduction}

The aim of speech enhancement (SE) is to remove background noise and improve the quality and intelligibility of noisy speech. SE has been widely applied in various fields, including hearing aids, text-to-speech (TTS) systems, and robust automatic speech recognition (ASR) \cite{eskimez2021human, iwamoto2022bad}.

Over the past decades, many SE algorithms have been developed \cite{mohammadiha2013supervised, xiang2021novel, wang2018supervised, luo2019conv, xiang2020parallel, hu20g_interspeech, wang2021compensation, zhao2022frcrn, zhao2024mossformer2}. Most current single-channel SE methods, whether statistical-based \cite{mohammadiha2013supervised, xiang2021novel}, traditional deep neural network-based \cite{wang2018supervised, zhao2022frcrn, zhao2024mossformer2, luo2019conv}, deep representation learning (DRL)-based \cite{xiang2022two, xiang2024deep}, or generative \cite{yang2024genhancer, lemercier2025diffusion}, mainly focus on directly estimating the full clean speech signal information (features), such as time-frequency (T-F) masks, waveforms \cite{luo2019conv}, spectrums \cite{hu20g_interspeech}, tokens \cite{yang2024genhancer, xue2024low}, or representations \cite{xiang2022two, xiang2024deep}. However, these methods often overlook the hierarchical semantic information.
%
When hierarchical semantic information is incorporated, these methods may more effectively handle complex acoustic environments, such as far-field or low signal-to-noise ratio (SNR) settings. In such scenarios, speech signals exhibit various intricate attributes—such as prosody, content, acoustics, and timbre—that, when preserved, contribute to a more accurate and robust speech enhancement process. This hierarchical structure enables a better understanding of these elements, improving the overall quality of the enhanced speech in challenging environments \cite{ju2024naturalspeech}. Moreover, integrating detailed speech attributes, such as prosody and content, into the estimation process enhances the preservation of speech features. This is especially beneficial for downstream tasks like ASR and TTS, which depend heavily on the fine-grained details of speech signals. By maintaining these attributes, the performance of these tasks is significantly boosted, enabling more precise and natural results in downstream applications \cite{eskimez2021human, iwamoto2022bad}.

Thus, this paper proposes a semantic information-based, step-by-step, SE (SISE) framework. This framework first predicts the semantic attribute directly from noisy speech and then uses this semantic attribute as a condition to predict the acoustic attribute for the final SE. Specifically, we first train a neural codec with factorized vector quantization to disentangle the speech waveform into subspaces corresponding to semantic and acoustic attributes. Next, we train a factorized semantic diffusion model to generate the semantic attribute of clean speech, using noisy speech as input. We then combine the estimated semantic attribute with the noisy input as a condition to train the acoustic diffusion model that predicts the acoustic attribute of clean speech. In our framework, both semantic and acoustic attributes are represented as corresponding tokens. Finally, during the inference stage, we utilize the diffusion model to generate the semantic and acoustic tokens of clean speech hierarchically, step by step. All estimated tokens are then passed to the codec's decoder to produce clean speech.

Compared to previous SE algorithms \cite{mohammadiha2013supervised, xiang2021novel, wang2018supervised, luo2019conv, xiang2020parallel, hu20g_interspeech, wang2021compensation, zhao2022frcrn, zhao2024mossformer2, xiang2022two, xiang2024deep, yang2024genhancer, lemercier2025diffusion}, which directly estimate the entire speech signal, the proposed method first predicts the semantic attribute of noisy speech. The semantic attribute typically contains less information than the full speech signal \cite{ju2024naturalspeech}, making it easier and more accurate to estimate in complex noisy environments. Once the semantic attribute is estimated, it serves as an additional condition to predict the more complex acoustic attribute, thereby reducing the difficulty of estimating acoustic information. Overall, our method models clean speech hierarchically, step by step, making estimation easier, more robust, and improving accuracy compared to existing SE methods. \cite{mohammadiha2013supervised, xiang2021novel, wang2018supervised, luo2019conv, xiang2020parallel, hu20g_interspeech, wang2021compensation, zhao2022frcrn, zhao2024mossformer2, xiang2022two, xiang2024deep, yang2024genhancer, lemercier2025diffusion}. Moreover, the proposed framework introduces the powerful discrete diffusion model \cite{lezama2022improved} as the generative model, ensuring the generation of high-quality speech. To our knowledge, this is the first work to analyze and disentangle the detailed semantic and acoustic attributes of speech in SE.


This paper is organized as follows. In Section 2, we present the details of the factorized codec and diffusion model. In Section 3, we describe the experimental settings and provide comparison results. Finally, we conclude the paper in Section 4.

\section{Semantic Information-Based SE}   

\begin{figure}[!tbp]
  \centering
  \includegraphics[width=\linewidth]{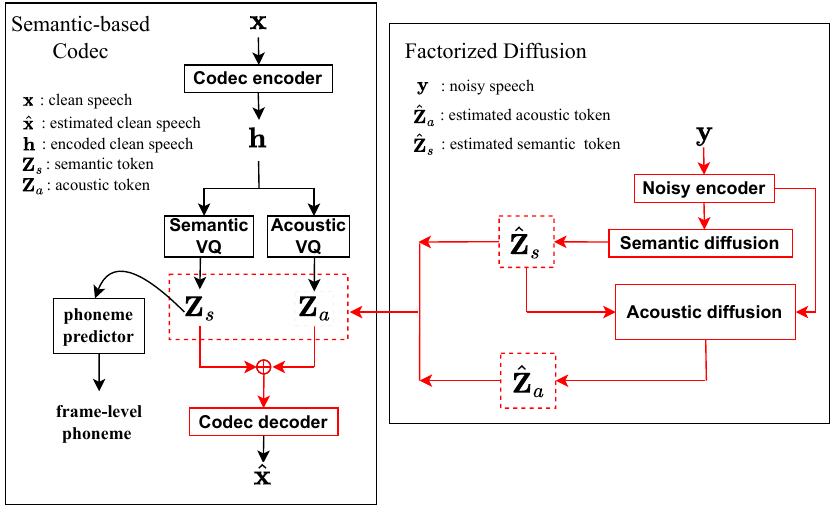}
  \caption{Overview of the SISE framework with a factorized codec and factorized diffusion model. The red part will be applied in both the training and inference processes, while the black part will be used only during the training stage.}
  \label{fig:SISE}
\end{figure}

In this section, we present the SISE framework. As shown in Fig.\ref{fig:SISE}, SISE consists of a semantic-based codec and a factorized diffusion model. In this work, we focus on achieving SE in far-field environments with complex noise. The observed signal is denoted as $\mathbf{y}$, and the clean speech signal is $\mathbf{x}$. To analyze the semantic information during SE, we factorize the speech into semantic and acoustic attributes. These attributes are represented by the corresponding tokens ${\mathbf{Z}}_{s}$ and ${\mathbf{Z}}_a$, respectively. The goal of the factorized diffusion model is to predict the factorized semantic and acoustic tokens given the noisy speech $\mathbf{y}$ as a condition. Specifically, a semantic diffusion model is used to estimate the semantic token given $\mathbf{y}$ as input. Then, an acoustic diffusion model estimates the acoustic token using the estimated semantic token ${\hat{\mathbf{Z}}_s}$ and $\mathbf{y}$ as conditions. Finally, the codec's decoder is applied to reconstruct the clean speech ${\hat{\mathbf{x}}}$ using the generated semantic token ${\hat{\mathbf{Z}}_s}$ and acoustic token ${\hat{\mathbf{Z}}_a}$. In Fig.\ref{fig:SISE}, the red part applies to both the training and inference processes, while the black part is used only during the training stage. We introduce the semantic-based codec in Section 2.1 and the factorized diffusion model in Section 2.2.

\subsection{Semantic-based codec}


To help SE to preserve semantic information in complex noisy environments, we design a semantic-based codec framework. This framework disentangles the clean speech waveform into subspaces representing semantic and acoustic attributes and then reconstructs the clean speech waveform from them.

As shown in Fig.~\ref{fig:SISE}, the semantic-based codec consists of a codec encoder, a semantic vector quantizer (VQ), an acoustic VQ, and a codec decoder. Given a clean speech input $\mathbf{x}$, we use the codec encoder to downsample the 16 kHz speech data by a downsample rate of 200 and obtain the encoded clean speech $\mathbf{h}$. The semantic and acoustic VQ is used to capture fine-grained speech attribute representations and produces discrete semantic tokens ${\mathbf{Z}}_s$ and acoustic tokens ${\mathbf{Z}}_a$, respectively. Specifically, to help the VQ disentangle semantic and acoustic information, we apply information bottleneck \cite{qian2020unsupervised, kumar2024high} and supervised learning \cite{ju2024naturalspeech} techniques. The information bottleneck forces the model to discard unnecessary information. The semantic and acoustic VQ apply the information bottleneck by projecting the encoder output into a low-dimensional space (i.e., 8 dimensions) and subsequently quantizing within this space. This ensures that each code embedding contains less information, thereby facilitating the disentanglement of semantic and acoustic attributes. After quantization, the quantized vectors are projected back to the original dimension. To further enhance the disentanglement, we also apply a supervised learning strategy. Since semantic tokens are strongly correlated with text or phonemes, the semantic token ${\mathbf{Z}}_s$ is sent to a phoneme predictor to estimate the frame-level phoneme labels. The frame-level phoneme labels are obtained using the Montreal Forced Aligner (MFA)\footnote{https://montreal-forced-aligner.readthedocs.io/en/latest/} toolkit. Based on our preliminary experiments, the information bottleneck and supervised learning are sufficient for effective disentanglement, and no additional techniques are needed. After obtaining the semantic token ${\mathbf{Z}}_s$ and acoustic token ${\mathbf{Z}}_a$, their sum ${\mathbf{Z}} = {\mathbf{Z}}_s + {\mathbf{Z}}_a$ is sent to the codec decoder to reconstruct the clean speech ${\hat{\mathbf{x}}}$.

The model architecture of the semantic-based codec follows \cite{kumar2024high}. Both the semantic and acoustic VQ use the residual vector quantization (RVQ) method \cite{zeghidour2021soundstream}. The semantic VQ consists of 1 quantizer (1 discrete token layer), while the acoustic VQ consists of 5 quantizers (5 discrete token layers). The codebook size for all quantizers is 1024, with a codebook dimension of 8, which is used to create the information bottleneck.

The discriminators and training losses of the codec also follow DAC \cite{kumar2024high}. However, due to the introduction of supervised learning, the total training loss ${\mathfrak{L}}_\text{total}$ for the generator differs slightly from that in \cite{kumar2024high}, and can be written as: 
\begin{align}
  {\mathfrak{L}}_\text{total} &= {\lambda_\text{rec}}{\mathfrak{L}}_\text{rec} + {\lambda_\text{adv}}{\mathfrak{L}}_\text{adv} +  \nonumber 
       {\lambda_\text{feat}}{\mathfrak{L}}_\text{feat}\\  
       & + {\lambda_\text{codebook}}{\mathfrak{L}}_\text{codebook} +  
       {\lambda_\text{commit}}{\mathfrak{L}}_\text{commit} + {\lambda_\text{sem}}{\mathfrak{L}}_\text{sem},
       \label{equation:codec}
\end{align}
where ${\mathfrak{L}}_\text{rec}$ is the multi-scale Mel-reconstruction loss \cite{kumar2024high}, and ${\mathfrak{L}}_\text{adv}$ includes both the multi-period discriminator and the multi-band multi-scale STFT discriminator, as proposed in \cite{kumar2024high}. Additionally, in Eqn.\eqref{equation:codec}, we also utilize the feature matching loss ${\mathfrak{L}}_\text{feat}$. For codebook learning, we use the codebook loss ${\mathfrak{L}}_\text{codebook}$ and the commitment loss ${\mathfrak{L}}_\text{commit}$ \cite{van2017neural}. ${\mathfrak{L}}_\text{sem}$ represents the phoneme prediction loss for supervised learning. The coefficients ${\lambda_\text{adv}}$, ${\lambda_\text{feat}}$, ${\lambda_\text{codebook}}$, ${\lambda_\text{commit}}$, ${\lambda_\text{rec}}$, and ${\lambda_\text{sem}}$ are used to balance each loss term. In our work, we set these coefficients as follows: ${\lambda_\text{adv}} = 4$, ${\lambda_\text{feat}} = 4$, ${\lambda_\text{codebook}} = 1$, ${\lambda_\text{commit}} = 1$, ${\lambda_\text{rec}} = 5$, and ${\lambda_\text{sem}} = 10$.

\subsection{Factorized diffusion}

\begin{figure}[!tbp]
  \centering
  \includegraphics[width=\linewidth]{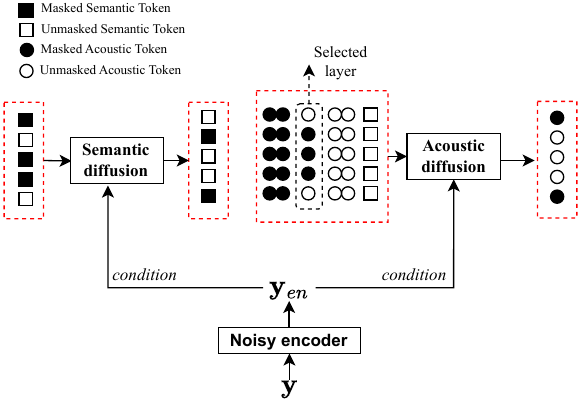}
  \caption{An overview of training diagram of factorized diffusion model, which consists of a semantic diffusion, an acoustic diffusion, and a noisy encoder.}
  \label{fig:diffusion}
\end{figure}

To model clean speech hierarchically, step by step, in complex noisy environments, we design a factorized diffusion framework consisting of a semantic diffusion, an acoustic diffusion, and a noisy encoder, as shown in Fig. \ref{fig:diffusion}. This section presents the details of the proposed factorized framework. 
Since semantic and acoustic diffusion use a similar training process for token generation, this section will focus on the acoustic diffusion training process and briefly explain the semantic one.

In Fig. \ref{fig:diffusion}, the noisy encoder downsamples the 16 kHz noisy speech by a downsample rate of 200 and obtains the encoded noisy speech ${\mathbf{y}}_\text{en}$, similar to the codec encoder. Our diffusion model design follows \cite{lezama2022improved}. For acoustic diffusion, the target discrete token sequence is ${\mathbf{Z}}_a = [{{z}}_{n,l}]_{n=1, l=1}^{N,L}$, ${N}$ is the total number of discrete token layers, and $L$ is the sequence length. In each training step, we randomly select the $j\,$th token layer to conduct the diffusion training, as shown in Fig.\ref{fig:diffusion}. The forward process at time $t$ is defined as masking a subset of tokens in ${\mathbf{z}}_n$ (where ${n}=j$) with the corresponding binary mask ${\mathbf{m}}_t=[{{m}}_{t,l}]_{l=1}^L$, formulated as ${\mathbf{z}}_{t,n}={\mathbf{z}}_{n} \odot {\mathbf{m}}_{t}$, where ${{z}}_{n,l}$ in ${\mathbf{z}}_{n}$ is replaced with a $[\texttt{Mask}]$ token if ${{m}}_{t,l}=1$, and left unmasked if ${{m}}_{t,l}=0$. Here, $m_{t,l} \sim \text{Bernoulli}(\sigma(t))$, with $\sigma(t) \in (0, 1]$ being a monotonically increasing function \cite{ju2024naturalspeech}. In this paper, we define $\sigma(t) = \sin\left({0.5\,\pi t}/{T}\right)$, where $t \in (0, T]$. Specifically, denote ${\mathbf{z}}_{0,n} = {\mathbf{z}}_{n}$ as the original token sequence and ${\mathbf{z}}_{T,n}$ as the fully masked sequence.

Regarding the reverse process, ${\mathbf{z}}_{0,n}$ is gradually restored by sampling from the reverse distribution $q({\mathbf{z}}_{t-\Delta t,n}|{\mathbf{z}}_{0,n}, {\mathbf{z}}_{t,n})$. Since ${\mathbf{z}}_{0,n}$ is unavailable during the inference stage, we utilize the diffusion model $p_{\theta}$, parameterized by $\theta$, to predict the masked tokens in the $j\,$th layer conditioned on $\mathbf{C}_a$, represented as $p_{\theta}({\mathbf{z}}_{0,n}| {\mathbf{z}}_{t,n}, {\mathbf{C}}_a)$. Regarding the acoustic diffusion model $p_{\theta}$, ${\mathbf{C}}_a = \{{\mathbf{y}}_\text{en}, \mathbf{Z}_{a\text{(unmasked)}}, \mathbf{Z}_s \}$, where ${\mathbf{Z}_{a\text{(unmasked)}}} = [{{z}}_{n,l}]_{n<j, l=1}^{N,L}$. Specially, if $n=1$, ${\mathbf{C}}_a = \{{\mathbf{y}}_\text{en}, \mathbf{Z}_s \}$. Finally, the input for the diffusion model $p_{\theta}$ is  $\{{\mathbf{z}}_{t,n}, \mathbf{Z}_{a\text{(masked)}}, {\mathbf{C}}_a\}$, where ${\mathbf{Z}_{a\text{(masked)}}}$ consists of tokens where $[{{z}}_{n,l}]_{n>j, l=1}^{N,L}$ in $\mathbf{Z}_a$ are replaced by the $[\texttt{Mask}]$ token. The input design of the acoustic diffusion follows Soundstorm \cite{borsos2023soundstorm}. The training target for $p_{\theta}$ is the masked tokens in the $j\,$th layer. The training diagram is shown in Fig.~\ref{fig:diffusion}. The parameters $\theta$ are optimized to minimize the negative log-likelihood of the masked tokens:
\begin{align}
  \mathcal{L}_{\text{mask(a)}} &= \mathbb{E}_{{\mathbf{Z}}_a \in \mathcal{D}, t \in [0,T]}  \nonumber \\  &\left[ - \sum_{l=1}^L m_{t,l}  \cdot \log(p_\theta({{z}}_{n,l} | {\mathbf{z}}_{t,n}, {\mathbf{C}}_a)) \right],
       \label{equation:diffusion_a}
\end{align}
where $\mathcal{D}$ represents the whole training dataset. The reverse transition distribution can be written as:
\begin{align}
  p({\mathbf{z}}_{t-\Delta t,n}|{\mathbf{z}}_{t,n}, {\mathbf{C}}_a)&= \mathbb{E}_{{\hat {\mathbf{z}}}_{0,n} \sim p_{\theta}({\mathbf{z}}_{0,n}| {\mathbf{z}}_{t,n}, {\mathbf{C}}_a)}  \nonumber \\  & \left[q({\mathbf{z}}_{t-\Delta t,n}|\hat{{\mathbf{z}}}_{0,n}, {\mathbf{z}}_{t,n}) \right].
       \label{equation:dis_a}
\end{align}

During the inference stage, the tokens are estimated layer by layer. The tokens estimated in the previous layer are used as conditions to estimate the tokens in the current layer, which is also similar to Soundstorm \cite{borsos2023soundstorm}. At each token layer, the masked tokens are progressively replaced, starting from the fully masked sequence ${\mathbf{z}}_{T,n}$, by iteratively sampling from $p({\mathbf{z}}_{t-\Delta t,n}|{\mathbf{z}}_{t,n}, {\mathbf{C}}_a)$. Following \cite{lezama2022improved, chung2022diffusion}, we first sample ${\hat {\mathbf{z}}}_{0,n}$ from $ p_{\theta}({\mathbf{z}}_{0,n}| {\mathbf{z}}_{t,n}, {\mathbf{C}}_a)$, and then sample ${\mathbf{z}}_{t-\Delta t,n}$ from $q({\mathbf{z}}_{t-\Delta t,n}|\hat{{\mathbf{z}}}_{0,n}, {\mathbf{z}}_{t,n})$. This involves remasking $\left\lfloor L \cdot \sigma\left(t - \Delta t\right) \right\rfloor$ (where $\left\lfloor \cdot \right\rfloor$ represents rounding down) tokens in ${\hat {\mathbf{z}}}_{0,n}$ with the lowest confidence scores. The confidence score of $\hat{{z}}_{n,l}$ in ${\hat {\mathbf{z}}}_{0,n}$ is defined as $ p_{\theta}({\hat{z}}_{l,n}| {\mathbf{z}}_{t,n}, {\mathbf{C}}_a)$ if ${{m}}_{t,l}=1$; otherwise, we set the confidence score of ${{z}}_{l,n}$ to 1. This means that tokens already unmasked in ${\mathbf{z}}_{t,n}$ will not be remarked.

For the semantic diffusion, there is a similar process. The difference is that it contains only one token layer and no additional conditions. Thus, its condition can be represented as: ${\mathbf{C}}_s = \{{\mathbf{y}}_\text{en}\}$, and the model details are also shown in Fig.~\ref{fig:diffusion}.

\section{Experimental Setup \& Results}
This section will evaluate the SE performance of the proposed algorithm. Specifically, we will first assess the enhanced speech quality in complex, far-field noisy environments. Then, we will evaluate its performance in a zero-shot TTS task. For the zero-shot TTS task, we will enhance the noisy prompt speech to determine whether the enhanced prompt speech improves the performance of the TTS system.


{\bf Datasets.} Our training dataset consists of three parts: clean speech, noise, and room impulse responses (RIRs). The clean speech includes English and Chinese data from Emilia \cite{he2024emilia}, each with 50K hours of speech (totaling 100K hours). The noise data is from the DNS Challenge 2021 corpus \cite{reddy2021interspeech}, which has about 150 audio classes and 60,000 clips. The RIRs are also selected from the DNS 2021 dataset, which provides 3,076 real and approximately 115,000 synthetic RIRs. The clean speech, noise, and RIRs are randomly mixed using the DNS script \cite{reddy2021interspeech}, with random SNR levels ranging from -10 dB to 20 dB. Finally, the noisy speech is obtained. For the semantic-based codec training, we only apply the clean speech data, which is processed by MFA to generate frame-level phoneme labels. For the training of the diffusion model, both clean and noisy speech data are utilized. All signals are down-sampled to 16 kHz. The evaluation dataset includes: (1) all real-world noisy recordings from the DNS Challenge 2021 test dataset, collected from various complex noisy environments; (2) SeedTTS test-en, a test set introduced in SeedTTS \cite{anastassiou2024seed} containing samples extracted from English public corpora, including 1,000 samples from the Common Voice dataset \cite{ardila2019common}; (3) SeedTTS test-zh, a test set introduced in SeedTTS containing samples extracted from Chinese public corpora, including 2,000 samples from the DiDiSpeech dataset \cite{guo2021didispeech}. The SeedTTS dataset is regarded as the clean speech dataset and also added with noise and RIRs for evaluation purposes, primarily simulating complex far-field noisy environments

{\bf Evaluation Metrics:} In this work, to evaluate the speech quality of the enhanced speech, we apply DNSMOS P.835 \cite{reddy2022dnsmos}. DNSMOS P.835 allows us to assess the speech quality (SIG), background noise quality (BAK), and overall quality (PMOS) of the audio samples. It has demonstrated a high degree of alignment with human ratings for evaluating speech quality. To evaluate whether the enhanced speech benefits the downstream zero-shot TTS task, we apply the Word Error Rate (WER) and speaker similarity (SIM-O) between the generated speech, using the enhanced speech as a prompt, and the clean prompt speech to evaluate TTS performance. For WER, we use Whisper-large-v3 \cite{radford2023robust} for SeedTTS test-en to transcribe English, and Paraformer-zh \cite{gao2023funasr} for SeedTTS test-zh to transcribe Chinese, following previous works \cite{anastassiou2024seed}. For SIM-O, the cosine similarity between the WavLM TDNN \cite{chen2022wavlm} speaker embedding of the generated samples and the clean prompt speech is used as the similarity score. For the downstream zero-shot TTS model, we use our trained TTS model, which has a structure similar to MaskGCT \cite{wang2024maskgct}, except that the speech semantic representation codec and speech acoustic codec in \cite{wang2024maskgct} are replaced by the proposed codec, which supports streaming TTS. This modification also results in a competitive TTS framework.

\begin{table}[!t]
 \centering
  \caption{{{Evaluation results for enhanced speech quality.}}}
  \label{tab: speech_quality_score}
  \centering
    \begin{tabular}{cccc}
    \toprule
    Method & BAK $\uparrow$ & SIG $\uparrow$  & PMOS $\uparrow$  \\
    \midrule
    \multicolumn{4}{c}{\textit{DNS 2021 test dataset}}   \\
    \midrule
    Noisy & 2.34 & 2.88 & 2.91 \\
    DCCRN \cite{hu20g_interspeech} & 3.68 & 3.05 & 3.39\\
    FRCRN \cite{zhao2022frcrn}  & 3.77   & 3.18  & 3.50 \\
    MOSS-FORMER \cite{zhao2024mossformer2}   & 3.88 & 3.15  & 3.43 \\
    SISE-w/o-dis  & {3.84}  & {3.25} & {3.53} \\
    proposed SISE  & \bf{3.88}  & \bf{3.32} & \bf{3.58} \\
    \midrule
    \multicolumn{4}{c}{\textit{SeedTTS test-en}}   \\
    \midrule
    Noisy &  2.25 & 2.05 & 2.62 \\
    DCCRN \cite{hu20g_interspeech} & 3.43 & 2.68 & 3.09 \\
    FRCRN \cite{zhao2022frcrn} &  3.74 & 2.80 & 3.21 \\
    MOSS-FORMER \cite{zhao2024mossformer2}  & 3.89 & 2.86 & 3.41 \\
    SISE-w/o-dis  & {3.90}  & {3.28} & {3.61} \\
    proposed SISE  & \bf{3.97} & \bf{3.47} & \bf{3.71} \\
     \midrule
    \multicolumn{4}{c}{\textit{SeedTTS test-zh}}   \\
    \midrule
    Noisy & 2.29  & 2.10 & 2.60 \\
    DCCRN \cite{hu20g_interspeech} & 3.45 & 2.71 & 3.13 \\
    FRCRN \cite{zhao2022frcrn} & 3.71  & 2.79 & 3.27 \\
    MOSS-FORMER \cite{zhao2024mossformer2}  & 3.76 & 2.82 & 3.38 \\
    SISE-w/o-dis  & {3.80}  & {3.40} & {3.62} \\
    proposed SISE  & \bf{4.02}& \bf{3.60} & \bf{3.73} \\

    \bottomrule       
  \end{tabular}
\end{table}

\begin{table}[!t]
 \centering
  \caption{{{Evaluation results for the Zero-Shot TTS system using enhanced speech as a prompt (Ground Truth refers to using clean speech as the prompt).}}}
  \label{tab: tts_eval}
  \centering
    \begin{tabular}{ccc}
    \toprule
    Method & \%WER $\downarrow$ & SIM-O $\uparrow$  \\
    \midrule
    \multicolumn{3}{c}{\textit{SeedTTS test-en}}   \\
    \midrule
    Ground Truth & 2.62  & 0.54  \\
    Noisy & 2.63  & 0.10  \\
    FRCRN \cite{zhao2022frcrn} & 2.60   & 0.15 \\
    MOSS-FORMER \cite{zhao2024mossformer2}  & 2.56  & 0.16 \\
    SISE-w/o-dis  & {2.57}  & {0.29} \\
    proposed SISE  & \bf{2.53}  & \bf{0.35}  \\
     \midrule
    \multicolumn{3}{c}{\textit{SeedTTS test-zh}}   \\
    \midrule
    Ground Truth & 1.93  & 0.71  \\
    Noisy & 2.20  & 0.25  \\
    FRCRN \cite{zhao2022frcrn} & 2.45   & 0.29 \\
    MOSS-FORMER \cite{zhao2024mossformer2}  & 2.47  & 0.30 \\
    SISE-w/o-dis  & {2.30}  & {0.41} \\
    proposed SISE  & \bf{2.19}  & \bf{0.53} \\

    \bottomrule       
  \end{tabular}
\end{table}

{\bf Implementation Details:} The codec is trained on eight NVIDIA H800 80GB GPUs. As described in \cite{kumar2024high}, its core component is a convolutional layer for upsampling or downsampling, followed by a residual block with Snake activations. The codec's encoder consists of 4 layers with downsampling rates [2, 2, 4, 5], while the codec's decoder has 4 layers with upsampling rates [5, 4, 2, 2]. The Adam optimizer is used for training with a learning rate of $2e-4$, $\beta_1 = 0.5$, and $\beta_2 = 0.9$. For more details on codec training, refer to \cite{kumar2024high}.

Similarly, our diffusion model is trained on eight NVIDIA H800 80GB GPUs. The noisy encoder follows the same structure as the codec's encoder. The Llama-style Transformer \cite{touvron2023llama} serves as the backbone for the whole diffusion model. The diffusion model uses an 8-layer Transformer with 8 attention heads and 1024 embedding dimensions, optimized by Adam with a learning rate of $1e-4$, $\beta_1 = 0.5$, and $\beta_2 = 0.9$.

In the inference stage, for semantic diffusion, we apply 15 steps by default for one semantic token layer. Sampling is done with top-k of 20 and temperature annealing from 1.5 to 0. For remasking, we add Gumbel noise to token confidences, as in \cite{wang2024maskgct}. For acoustic diffusion, the default steps are $[10, 1, 1, 1, 1]$ for the 5 acoustic token layers. The sampling strategy is the same as for semantic diffusion, except greedy sampling is used when the inference step is 1.

{\bf Baselines:} To evaluate SE performance, the proposed SISE framework is compared with state-of-the-art (SOTA) SE models: FRCRN \cite{zhao2022frcrn}, MOSS-FORMER \cite{zhao2024mossformer2}, and DCCRN \cite{hu20g_interspeech}. FRCRN and MOSS-FORMER are implemented using ClearerVoice-Studio\footnote{https://github.com/modelscope/ClearerVoice-Studio/tree/main}, while DCCRN is implemented with the open-source code\footnote{https://github.com/huyanxin/DeepComplexCRN}. All three\cite{zhao2022frcrn, zhao2024mossformer2, hu20g_interspeech} are competitive SOTA SE algorithms. Due to their superior performance, MOSS-FORMER and FRCRN are also used as reference methods in the evaluation of SE performance for the downstream zero-shot TTS task. Additionally, to verify the effectiveness of step-by-step attribute prediction, we conduct an ablation study (SISE-w/o-dis), which uses DAC \cite{kumar2024high} as the codec without semantic attribute disentanglement, relying only on a diffusion model to predict all tokens.

{\bf Experimental Results:} Table~\ref{tab: speech_quality_score} presents the evaluation results for enhanced speech quality. The proposed SISE framework achieves the highest BAK, SIG, and PMOS scores across all three test datasets. Additionally, codec + diffusion-based methods (SISE-w/o-dis and SISE) significantly outperform traditional SOTA SE algorithms, especially in complex far-field noisy environments (SeedTTS test-en and test-cn), with PMOS improvements exceeding 0.3. Comparing SISE-w/o-dis and SISE, we observe a clear PMOS improvement with SISE, indicating that hierarchical modeling of clean speech step by step further enhances SE performance. Table~\ref{tab: tts_eval} shows the evaluation results for the Zero-Shot TTS System using enhanced speech as a prompt. The results reveal that enhanced prompts from all algorithms do not affect TTS system robustness, with similar WER scores. However, SISE significantly improves speaker similarity, demonstrating its benefits for downstream TTS tasks.

\section{Conclusions}
This paper presents a novel SE method that leverages semantic information-based factorization. Unlike traditional SE algorithms, our approach hierarchically models both semantic and acoustic attributes, improving the estimation of clean speech in complex acoustic environments. Experimental results show that the proposed SISE outperforms current SOTA algorithms, particularly in far-field noisy conditions, highlighting the effectiveness of semantic attribute estimation in SE. Furthermore, SISE demonstrates improvements for downstream TTS tasks. While this work primarily focuses on SE, our method is also applicable to other speech recovery tasks, such as clipping and bandwidth limitation distortion. Future work will include a more comprehensive speech quality evaluation of these applications.


\ninept 
\scriptsize 
\bibliographystyle{IEEEtran}
\bibliography{IEEEabrv,myabrv_new,my_reference,mybib}

\end{document}